\documentclass[12pt,journal,compsoc]{IEEEtran}
\ifCLASSOPTIONcompsoc
\else
\fi
\ifCLASSINFOpdf
\else
\fi
\usepackage[cmex10]{amsmath}
\usepackage{amsfonts}
\usepackage{multirow}
\usepackage{mathrsfs}
\usepackage{amsmath}
\usepackage{amssymb}
\usepackage{subfigure}
\usepackage{amsmath}
\usepackage{pgf,tikz,pgfplots}
\usepackage{hyperref}

\usepackage{soul}
\usepackage{color}

\newtheorem{theorem}{Theorem}

\newtheorem{corollary}{Corollary}

\newcommand\BibTeX{{\rmfamily B\kern-.05em \textsc{i\kern-.025em b}\kern-.08em
T\kern-.1667em\lower.7ex\hbox{E}\kern-.125emX}}
\newcommand{\diag}{\mathop{\mathrm{diag}}\nolimits}

\newcounter{remark}
\newenvironment{remark}{\refstepcounter{remark}\vspace{1ex}
{\sc Remark \theremark.}\hspace{0.3em}\parindent=0pt}{\vspace{1ex}}

\usepackage[ruled]{algorithm2e}

\usepackage{algorithmic}

\usepackage{graphicx}

\usepackage{array}

\begin{document}
%
\title{A parallel structured divide-and-conquer algorithm for symmetric tridiagonal eigenvalue problems}
%

%
%

\author{Xia~Liao,~Shengguo~Li,~Yutong~Lu
  and~Jose E. Roman 
\IEEEcompsocitemizethanks{\IEEEcompsocthanksitem X. Liao and S. Li are with the College
of Computer Science, National University of Defense Technology, Changsha, China.\protect\\
E-mail: nudtlsg@gmail.com
\IEEEcompsocthanksitem Y. Lu is with the National Supercomputer Center
in Guangzhou, and the School of Data and Computer Science, Sun Yatsen University, Guangzhou, China, 510006.%
\IEEEcompsocthanksitem J.~E. Roman is with the D. Sistemes Inform\`{a}tics i Computaci\'{o}, Universitat Polit\`{e}cnica de Val\`{e}ncia, Cam\'{i} de Vera s/n, 46022 Val\`{e}ncia, Spain.} 
\thanks{Published on  IEEE Trans. Parallel and Distributed Systems. DOI: \url{https://doi.org/10.1109/TPDS.2020.3019471}. Please cite the journal version.}}

%
%

\markboth{Journal of \LaTeX\ Class Files,~Vol.~11, No.~4, December~2012}%
{Shell \MakeLowercase{\textit{et al.}}: Bare Demo of IEEEtran.cls for Computer Society Journals}
%



\IEEEtitleabstractindextext{%
\begin{abstract}
  In this paper, a parallel structured divide-and-conquer (PSDC) eigensolver is proposed for symmetric
  tridiagonal matrices based on ScaLAPACK and a parallel structured matrix multiplication
  algorithm, called PSMMA.
  Computing the eigenvectors via matrix-matrix multiplications is the most computationally expensive part of the divide-and-conquer algorithm,
  and one of the matrices involved in such multiplications is a rank-structured Cauchy-like matrix.
  By exploiting this particular property, PSMMA constructs the local matrices by using
  generators of Cauchy-like matrices without any communication,
  and further reduces the computation costs by using a structured low-rank approximation algorithm.
  Thus, both the communication and computation costs are reduced.
  Experimental results show that both PSMMA and PSDC are highly scalable and scale to 4096 processes at least.
  PSDC has better scalability than PHDC that was proposed in [J. Comput. Appl. Math. 344 (2018) 512--520] and
  only scaled to 300 processes for the same matrices.
  Comparing with \texttt{PDSTEDC} in ScaLAPACK, PSDC is always faster and achieves $1.4$x--$1.6$x speedup for some matrices
  with few deflations.
  PSDC is also comparable with ELPA, with PSDC being faster than ELPA when using few processes and
  a little slower when using many processes.

\end{abstract}

}

\maketitle

\IEEEdisplaynontitleabstractindextext

%
\IEEEpeerreviewmaketitle

\section{Introduction}

Computing the eigendecomposition of a symmetric tridiagonal matrix is an important linear
algebra problem and is widely used in many fields of science and
engineering.
It is usually solved by divide-and-conquer (DC)~\cite{Cuppen81,Gu-eigenvalue}, QR~\cite{Golub-book2}, MRRR~\cite{MRRR-LAA,Petschow-sisc}, and
some other methods.
The DC algorithm is now the default method in LAPACK~\cite{anderson1999lapack} and
ScaLAPACK~\cite{Scalapack} when the eigenvectors are required.
DC is usually very efficient in practice, requiring only $O(n^{2.3})$ flops for matrices with
dimension $n$ on average~\cite{Tisseur-DC},
but its complexity can still be $O(n^3)$ for some matrices with few deflations.
The performance of an algorithm not only depends on the number of floating point operations,
but also other ingredients such as communication and data movements, etc.
Some communication-avoiding algorithms have been developed recently~\cite{BDHS11,Hoemmen-sisc}.
In this work, we aim to accelerate the parallel DC algorithm on distributed memory machines
by reducing both the \emph{computation} complexity and \emph{communication} complexity.
Our algorithm can be much faster than the ScaLAPACK routine \texttt{PDSTEDC} for those difficult matrices,
and can scale to thousands of processes.
Experimental results are included in section~\ref{sec:num}.

It is known that some Cauchy-like matrices with off-diagonally low-rank properties appear in the
DC algorithm~\cite{Cuppen81,Gu-eigenvalue},
which can be approximated by hierarchically semiseparable (HSS) matrices~\cite{ChandrasekaranGu04,ChandrasekaranDe06}.
Then, the worst case complexity of DC can be reduced from $O(n^3)$ to $O(n^2 r)$, where $r$ is a modest number and is usually much smaller than a large $n$, see~\cite{LSG-NLAA}.
This technique was extended for the bidiagonal and banded DC algorithms for the SVD problem on a
shared memory multicore platform in~\cite{Shengguo-SIMAX2,Liao-camwa}.
For distributed memory machines, a parallel DC algorithm is similarly
proposed in~\cite{li2018efficient} by using STRUMPACK (STRUctured Matrices PACKage)~\cite{Strumpack},
which provides some distributed parallel HSS algorithms.
The accelerated DC proposed in~\cite{li2018efficient} was called \emph{PHDC}.
However, numerical results show that for the simple matrix-matrix multiplication operations,
STRUMPACK is not as scalable as \texttt{PDGEMM},
and may become slower than \texttt{PDGEMM} when using 300 or more processes on Tianhe-2 supercomputer.
See~\cite{li2018efficient} for details.

Instead of using HSS, this work exploits a much simpler type of
rank-structured matrix, called BLR (Block low-rank format~\cite{amestoy2015improving}).
Compared with HSS, BLR abandons the hierarchy but compresses the off-diagonal
blocks. It loses the near-linear complexity of other hierarchical matrices such as
$\mathcal{H}$-matrix~\cite{Hackbusch1999}, $\mathcal{H}^2$-matrix~\cite{Hackbusch-Sauter2000}, and HSS matrix.
Because of its simple structure, BLR is easy to implement in parallel, and
often improves the scalability of corresponding algorithms,
see~\cite{amestoy2015improving,yamazaki2019distributed-memory} for more details.

In this paper, we propose a parallel structured matrix-matrix multiplication
algorithm for Cauchy-like matrices, which will be named \emph{PSMMA}.
It exploits the off-diagonal low-rank property of matrices like BLR,
and it can further reduce the communication cost by constructing local submatrices
using the \emph{generators}, which will be explained in section~\ref{sec:algorithm}.
Our main contributions include the following:
\begin{itemize}
  \item We propose a parallel structured matrix multiplication algorithm (PSMMA) for structured matrices including
        Cauchy-like, Toeplitz, Hankel, Vandermonde, etc.
        PSMMA can reduce both the communication and computation costs by using low-rank approximations.
        To the best of the authors' knowledge, none of the matrix multiplication algorithms has been developed
        to reduce the communication cost by exploiting the structure of matrices.

  \item PSMMA works for matrices both in the block cyclic data distribution (BCDD) form (like ScaLAPACK) and
        block data distribution (BDD) form (2D block partitioning).
        It also works for general process grids.

  \item Combining PSMMA with the DC algorithm in ScaLAPACK, we propose a parallel structured
        DC algorithm (PSDC), which can be much faster than \texttt{PDSTEDC} in ScaLAPACK.
        PSDC is also competitive with ELPA~\cite{Elpa}.


\end{itemize}

The process of PSMMA is similar to Cannon~\cite{Cannon_MM} and Fox~\cite{fox1987matrix} algorithms.
However, PSMMA works for matrices in BCDD form and works for any rectangular process grids.
From this perspective, PSMMA is more like PUMMA~\cite{choi1994pumma}, a generalized Fox algorithm.
PSMMA is more efficient than PUMMA for structured matrices and details are shown in
section~\ref{sec:comparison}.
It has three advantages compared with PUMMA.
One advantage is that PSMMA constructs the required submatrix locally by using the generators without
communication and thus requires less \emph{communication}.
Another one is that PSMMA combines with low-rank approximations and therefore
the \emph{computation complexity} is also reduced.
The third one is that PSMMA requires less workspace
and the size of local matrix multiplications is also larger than PUMMA.
In this paper, SRRSC~\cite{Shengguo-SIMAX2,GX-Toeplitz} is used to
compute the low-rank approximations of Cauchy-like matrices in PSMMA, which only
requires linear storage.
Note that \texttt{PDGEMM} implements an algorithm similar to SUMMA~\cite{1997summa}.
Compared with SUMMA~\cite{1997summa}, which is based on the outer product form of matrix multiplication,
PSMMA can naturally exploit the off-diagonal
low-rank property of matrices.

By incorporating PSMMA into the DC algorithm in ScaLAPACK~\cite{Tisseur-DC},
we obtain a highly scalable DC algorithm, which has much better
scalability than the previous PHDC algorithm~\cite{li2018efficient}.
To distinguish from PHDC, we call the newly proposed algorithm
\emph{parallel structured DC} algorithm (PSDC).
Numerical results show that PSDC is always faster than \texttt{PDSTEDC} in ScaLAPACK, and scales to
4096 processes at least.
That is because PSDC requires both less computations and communications than \texttt{PDSTEDC}.
The speedups of PSDC over \texttt{PDSTEDC} can be up to $1.4$x-$1.6$x for some matrices with
dimension $30,000$ on Tianhe-2 supercomputer.
Note that PHDC in~\cite{li2018efficient} can only scale to 300 processes for the same matrices.

The remaining sections of this paper are organized as follows.
Section~\ref{sec:prelim} introduces the DC algorithm,
the SRRSC algorithm for
constructing low-rank approximations of Cauchy-like matrices, and
some classical parallel matrix multiplication algorithms.
Section~\ref{sec:algorithm} presents the newly proposed
parallel structured matrix multiplication algorithm PSMMA, and describes
the implementation details of PSDC, which combines PSMMA with
the parallel tridiagonal DC algorithm in ScaLAPACK.
All the experimental results are reported in section~\ref{sec:num}, and
some future works are included in section~\ref{sec:future}.
Conclusions are drawn in section~\ref{sec:conclusion}.

\section{Preliminaries}
\label{sec:prelim}

Assume $T$ is a symmetric tridiagonal matrix,
\begin{equation}
\label{eq:t}
T  =
\left [
\begin{array}{cccc}
a_{1}& b_{1} &  &   \\
b_{2}& \ddots & \ddots &  \\
 & \ddots  & a_{n-1} & b_{n-1} \\
 & & b_{n-1}  & a_{n}
\end{array}
\right ]. \\
\end{equation}
We briefly introduce some formulae of Cuppen's divide-and-conquer
algorithm~\cite{Cuppen81,Tisseur-DC}.
The ScaLAPACK routine also implements this version of DC algorithm~\cite{Tisseur-DC}
based on rank-one update.

\begin{algorithm}[htb]
\label{alg:dc}
\SetAlgoNoLine
\KwIn{$T\in\mathbb{R}^{n\times n}$}
\KwOut{eigenvalues $\Lambda$, eigenvectors $Q$}
\eIf{ the size of $T$ is small enough}{
apply the QR algorithm and compute $T = Q \Lambda Q^T $ \;
\Return $Q$ and $\Lambda$\;
}{
form $T= \begin{bmatrix} T_1 & \\ & T_2 \end{bmatrix}+b_k vv^T$\;
\textbf{call} DC($T_1,Q_1,\Lambda_1$)\;
\textbf{call} DC($T_2,Q_2,\Lambda_2$)\;
form $M = D+b_kuu^T$\ from $Q_i, \Lambda_i$ and $v$ ($i=1,2$), where $D=\diag(\Lambda_1,\Lambda_2)$\;
find eigenvalues $\Lambda$ and eigenvectors $\widehat{Q}$ of $M$\;
compute the eigenvectors of $T$ as $ Q= \begin{bmatrix} Q_1 & \\ & Q_2 \end{bmatrix} \widehat{Q}$\;
    \Return $Q$ and $\Lambda$\;
 }
\caption{DC($T, Q, \Lambda$) algorithm for computing the eigendecomposition of a symmetric tridiagonal matrix.}
\end{algorithm}

Firstly, $T$ is decomposed into the sum of two matrices,
\begin{equation}
  \label{eq:T2}
  T=
  \begin{bmatrix}
    T_1 & \\ & T_2
  \end{bmatrix}+b_k vv^T,
\end{equation}
where $T_1\in \mathbb{R}^{k\times k}$, $b_k$ is the off-diagonal element at
the $k$th row of $T$ and $v=[0,\ldots,1,1,\ldots,0]^T$ with
ones at the $k$th and $(k+1)$th entries.
If $T_1=Q_1 \Lambda_1Q_1^T$ and $T_2=Q_2 \Lambda_2 Q_2^T$, then $T$ can be written as
\begin{equation}
  \label{eq:T3}
  T=
  \begin{bmatrix}
    Q_1 & \\ & Q_2
  \end{bmatrix} \left(
  \begin{bmatrix}
    \Lambda_1 & \\ & \Lambda_2
  \end{bmatrix} + b_k uu^T \right)
\begin{bmatrix}
  Q_1^T & \\ & Q_2^T
\end{bmatrix},
\end{equation}
where $u=
\begin{bmatrix}
  Q_1^T & \\ & Q_2^T
\end{bmatrix} v =
\begin{bmatrix}
  \text{last col. of } Q_1^T \\
  \text{first col. of } Q_2^T
\end{bmatrix}.
$

Since $Q_1$ and $Q_2$ are orthogonal matrices, the problem is reduced to computing the
spectral decomposition of a diagonal plus rank-one matrix,
\begin{equation}
\label{eq:rankone}
M\equiv D+b_{k} uu^T = \widehat{Q} \Lambda \widehat{Q}^T,
\end{equation}
where 
$D=\diag(\Lambda_1,\Lambda_2)$ is a diagonal matrix,
$\Lambda$ is a diagonal matrix whose
diagonal elements are the eigenvalues of matrix $M$,
and $\widehat{Q}$ is the eigenvector matrix of $M$.
Then, the eigenvector matrix of $T$ is computed as
\begin{equation}
\label{eq:updateV}
\left [
\begin{array}{cc}
Q_{1} &   \\
& Q_{2}
\end{array}
\right ] \\
\widehat{Q}.
\end{equation}
The eigenvalues $\lambda_{i}$ of $ D +b_{k} uu^T$ are the roots of the secular equation
\begin{equation}
\label{eq:secular}
f(\lambda)=1+ b_{k} \frac {u_{k}^2 }{d_{k}-\lambda}=0,
\end{equation}
where $d_k$ is the $k$th diagonal entry of $D$, $u_k$ is the $k$th component of $u$.
Then, the eigenvector is computed as
\begin{equation}
\label{eq:eigenvector}
\hat{q_{i}}=(D-\lambda_{i}I)^{-1}u.
\end{equation}
%

The main observation of works~\cite{li2018efficient,LSG-NLAA} is that
\begin{equation}
\label{eq:Q_Cauchy}
\widehat{Q}=\left( \frac{u_i v_j}{d_i-\lambda_j} \right)_{i,j},
\end{equation}
where $v_j=1/\sqrt{\sum_{k=1}^n \frac{u_k^2}{(d_k-\lambda_j)^2}}$, is a Cauchy-like matrix,
and the vectors $u, v\in \mathbb{R}^n$ and $d=(d_1,\cdots, d_n)^T, \lambda=(\lambda_1,\cdots,\lambda_n)^T$
are called \emph{generators}.

The whole classical DC algorithm is shown in Algorithm~\ref{alg:dc}.
The main computational task of DC lies in computing the eigenvectors via matrix-matrix multiplications (MMM)~\eqref{eq:updateV},
which costs $O(n^3)$ flops. 
Since $\widehat{Q}$ is a Cauchy-like matrix and off-diagonally low-rank,
MMM~\eqref{eq:updateV} can be accelerated by using HSS matrix algorithms,
and the computational complexity can be reduced significantly, see~\cite{li2018efficient,LSG-NLAA}
for more details.
The aim of this work is not only to reduce the computation cost of MMM~\eqref{eq:updateV}
but also its communication cost in the distributed memory environment.
For simplicity, we do not consider \emph{deflation} in~\eqref{eq:updateV}.
About the \emph{deflation} process, we refer the interested readers to~\cite{Cuppen81,Gu-rank1} and section~\ref{sec:deflation}.

\subsection{SRRSC low-rank approximation}
\label{sec:srrsc}

A novel low-rank approximation method for Cauchy-like matrix is proposed in~\cite{Shengguo-SIMAX2,GX-Toeplitz}, which
only requires linear storage.
For completeness, this method is introduced briefly in this section,
which only works on the generators.

Assume that $A$ is an $n\times n$ Cauchy-like matrix, $A=(\frac{u_iv_j}{d_i-w_j})_{i,j}$.
The following factorization is called the $k$th \emph{Schur complement factorization} of $A$,
\begin{equation}
  \label{eq:col-schur}
  A=
  \begin{bmatrix}
    A_{11} & A_{12} \\ A_{21} & A_{22}
  \end{bmatrix}
  =
  \begin{bmatrix}
    A_{11} & \\ A_{21} & A^{(k)}
  \end{bmatrix}
  \begin{bmatrix}
    I & Z^{(k)} \\ & I
  \end{bmatrix},
\end{equation}
where $A_{11} \in \mathbb{R}^{k\times k}$ and $A^{(k)}$ is called the $k$th Schur complement.

One good property of a Cauchy-like matrix is that its $k$th Schur complement $A^{(k)}$ is also Cauchy-like~\cite{GKO,Gu-Cauchy}, and
its generators can be computed recursively as follows.

\begin{theorem}
\label{thm:uvk}
The $k$th Schur complement $A^{(k)}$ satisfies
\begin{equation*}
\label{eq:sckk}
D_k A^{(k)} - A^{(k)} W_k = u^{(k)}(k+1:n)\cdotp v^{(k)T}(k+1:n),
\end{equation*}
with $D_{k+1} = \diag(d_{k+1},\cdots, d_n)$ and $W_{k+1} = \diag(w_{k+1},\ldots, w_n)$.
Then
\begin{equation}
\label{eq:UU}
u^{(k)}(k+\ell) = u^{(k-1)}(k+\ell)\cdotp \frac{d_{k+\ell} - d_{k}}{d_{k+\ell} - w_{k}},
\end{equation}
\begin{equation}
\label{eq:VV2}
v^{(k)}(k+\ell) = v^{(k-1)}(k+\ell) \cdotp \frac{w_{k+\ell}-w_{k}}{w_{k+\ell}-d_{k}},
\end{equation}
with $1\le \ell \le n-k, k\ge 1$, $A^{(0)}=A$, $u^{(0)}=u$ and $v^{(0)}=v$.
\end{theorem}


\begin{corollary}
\label{cor:uv}
For $1\le \ell \le n-k, k\ge 1$, the generators of $A^{(k)}$ are
\begin{equation}
  \label{eq:u-recursive}
u^{(k)}(k+\ell)  = \prod_{j=1}^k \frac{d_{k+\ell}-d_j}{d_{k+\ell}-w_j}\cdotp u^{(0)}(k+\ell), \\
\end{equation}
\begin{equation}
  \label{eq:v-recursive}
v^{(k)}(k+\ell)  = \prod_{j=1}^k \frac{w_{k+\ell}-w_j}{w_{k+\ell}-d_j}\cdotp v^{(0)}(k+\ell). \quad \square
\end{equation}
\end{corollary}

It is easy to prove that $Z^{(k)}$ is also Cauchy-like, and
its generators can also be computed recursively.
\begin{theorem}
\label{thm:Nkk}
The generators of $Z^{(k)}$ satisfy the displacement equation
\begin{equation*}
\label{eq:Zk-gen}
W_1^{(k)} Z^{(k)} -Z^{(k)} W_2^{(k)} = y^{(k)}(1:k) \cdotp v^{(k)}(k+1:n)^T,
\end{equation*}
where $W_1^{(k)} = \diag(w_1,\ldots, w_k)$ and $W_2^{(k)}=\diag(w_{k+1},\ldots, w_n)$, and
  \begin{equation}
    \label{eq:yk}
    y^{(k)}(i) = \prod_{1\le j \le k, j\ne i} \frac{d_j-w_i}{w_j-w_i} \cdotp \frac{d_i-w_i}{v^{(0)}(i)},
  \end{equation}
and $v^{(k)}(k+1:n)$ are computed recursively by equation~\eqref{eq:v-recursive}.
Furthermore,
\begin{equation*}
\label{eq:y}
y^{(k)}(i) = \begin{cases} y^{(k-1)}(i) \cdotp \frac{d_{k}-w_i}{w_{k} - w_i}, & \text{if  $1\le i \le k-1$}, \\
    \prod_{j=1}^{k-1} \frac{w_{k} - d_j}{w_{k} - w_j} \cdotp \frac{d_{k}-w_{k}}{v^{(0)}(k)}, & \text{if $i = k$}.  \quad \square \end{cases}
\end{equation*}

\end{theorem}

%

Since permutation does not destroy the structure of Cauchy-like matrices, we can permute generators
$u^{(k)}, v^{(k)}, d$ and $w$ to make the first entry of
$A^{(k)}$, $A^{(k)}(1,1)=\frac{u^{(k)}(k+1)v^{(k)}(k+1)}{d_{k+1}-w_{k+1}}$, large.
A complete pivoting strategy could be used.
Some efficient pivoting strategies have been proposed in~\cite{Gu-Cauchy} and~\cite{Pan00}.
We used the pivoting strategy proposed in~\cite{Shengguo-SIMAX2}.

If the entries of $A^{(k)}$ are negligible, and then by ignoring $A^{(k)}$, we get a low-rank approximation to $A$,
\begin{equation}
  \label{eq:sid}
  AP\approx
  A(1:n,\mathscr{T})
  \begin{bmatrix}
    I & Z^{(k)}
  \end{bmatrix},
\end{equation}
where $\mathscr{T}=\{i_1,i_2,\ldots,i_k\}\subset \{1,2,\ldots, n\}$ and
$P$ is a permutation matrix which records the column permutations during the pivotings.
To be more specific, $A(1:n,\mathscr{T})$ consists of a subset of columns of $A$, and
from Theorem~\ref{thm:Nkk},
\begin{equation}
\label{eq:geneZ}
Z^{(k)}_{ij} = \frac{u^{(k)}(k+i)y^{(k)}(j)}{w(i)-w(k+j)},
\end{equation}
where $w$ is the array after permutation.

\subsection{Parallel Matrix-Matrix multiplications}
\label{sec:pmmm}

Matrix-matrix multiplication is a very important computational kernel of
many scientific applications.
In this subsection, we briefly introduce some parallel matrix multiplication
algorithms, which compute $C=A\times B$.

Cannon algorithm~\cite{Cannon_MM} was the first efficient
algorithm for parallel matrix multiplication providing
theoretically optimal communication cost.
However it requires the process grid to be square, which
limits its practical usage.
Fox algorithm proposed in~\cite{fox1987matrix} has
the same problem.
The PUMMA algorithm~\cite{choi1994pumma} is a generalized
Fox algorithm, and it works for a general $P\times Q$ processor grid.
PUMMA was designed for ScaLAPACK and used the BCDD form.

The current version of ScaLAPACK implements SUMMA~\cite{1997summa}, which was proposed
in the mid-1990s and designed for a general processor grid.
It also uses the block-cyclic data distribution form.
It implements the outer product form of matrix multiplication, and
allows to pipeline them.
PUMMA implements the inner product form, and requires the largest possible matrices
for computations and communications.
Some more efficient matrix multiplication algorithms have been proposed recently,
like 2.5D algorithm~\cite{Solomonik_2p5mm}, CARMA~\cite{demmel2013communication},
CTF~\cite{solomonik2013cyclops},
and COSMA~\cite{kwasniewski2019red}, and many others.
Since they are not quite related to this current work,
we do not attempt to give a complete literature review.

PUMMA is more appropriate for the rank-structured matrices than SUMMA, since
the large off-diagonal blocks can be compressed by low-rank approximations.
So are Cannon and Fox algorithms.
It is not obvious for SUMMA to exploit the rank-structured property of the
input matrices.

In this paper, we propose a parallel structured matrix multiplication algorithm,
which is called PSMMA for short.
We assume at least one of the two matrices is a structured matrix.
By 'structured matrices' we mean those matrices which can be expressed using $O(n)$ parameters where
$n$ is the dimension of matrix, e.g. Cauchy-like, Toeplitz, Hankel, and Vandermonde matrices~\cite{Pan-book}.
PSMMA can be based on the BCDD structure like PUMMA and ScaLAPACK,
and it can also be based on the BDD structure like Cannon or Fox algorithms.
As shown later, the second approach is more efficient to exploit the off-diagonal low-rank structure.
Its drawback is that it requires to redistribute the matrix from the BCDD form to BDD form, since
the matrices are initially stored in BCDD form in ScaLAPACK routines.
The PSMMA in BCDD form fits well with ScaLAPACK routines and
does not need any data redistribution.

By exploiting the special structure of matrix, PSMMA can reduce
both the computation and communication costs.
To the best of the authors' knowledge, this work is the first one
proposing a reduction of the communication cost
of matrix multiplication algorithms by exploiting the structures of matrices.

\section{Algorithm Proposed}
\label{sec:algorithm}

The main contributions of this paper consist of two parts.
First, we design a parallel structured matrix multiplication algorithm for structured matrices
in section~\ref{sec:structuredFox}, which is called PSMMA
and can reduce both the computation and communication costs.

Secondly, section~\ref{sec:deflation} shows how to combine PSMMA with the parallel tridiagonal DC algorithm
in ScaLAPACK.
It illustrates how to modify several routines in ScaLAPACK in order to use PSMMA.
The parallel structured DC algorithm is called PSDC, and its whole
procedure is summarized in Algorithm~\ref{alg:psdc}.
A cartoon is included in Fig.~\ref{fig:psdcfig} to show the whole workflow of PSDC.

\subsection{PSMMA for structured matrices}
\label{sec:structuredFox}

In this subsection, we introduce PSMMA to compute $C=A\times B$, where $A\in \mathbb{R}^{m\times k}$ is a general matrix,
$B\in \mathbb{R}^{k\times n}$ is a structured matrix, and its entries can be represented
by $O(n+k)$ parameters.
The case that $A$ is a structured matrix and $B$ is a general matrix is similar.
When both $A$ and $B$ are structured matrices, all operations can be performed locally
without any communication, which will be explained later.
In the following sections, we assume $A$ is stored in BCDD form and $B$ is represented by
its generators.

\begin{figure}[ptbh]
\centering
\subfigure[Matrix point-of-view.]{
\includegraphics[width=2.5in,height=2.5in]{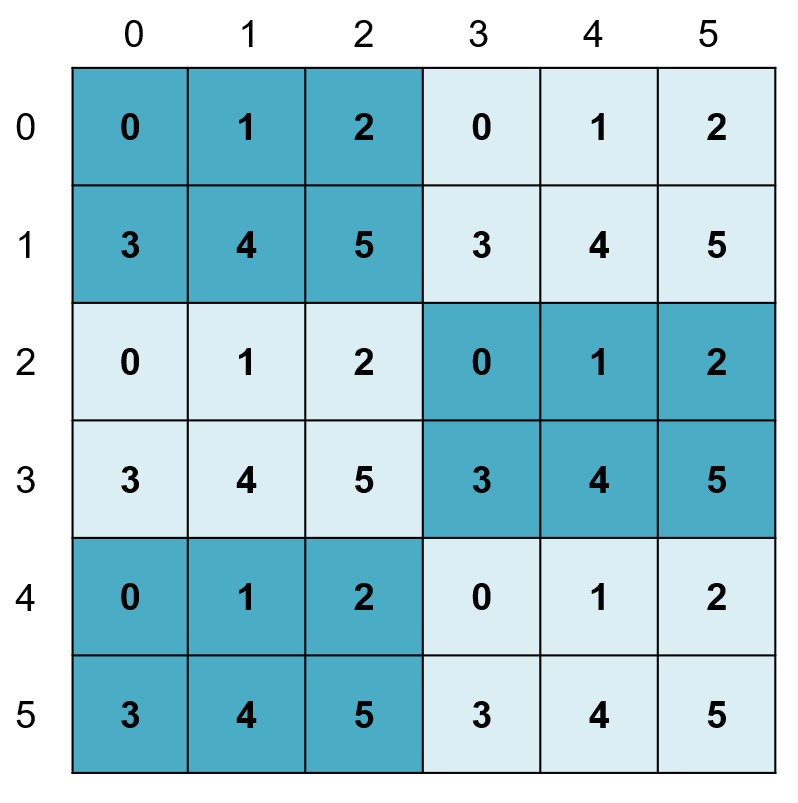}
\label{fig:cyclic}}
\subfigure[Process point-of-view.]{
\includegraphics[width=2.5in,height=2.5in]{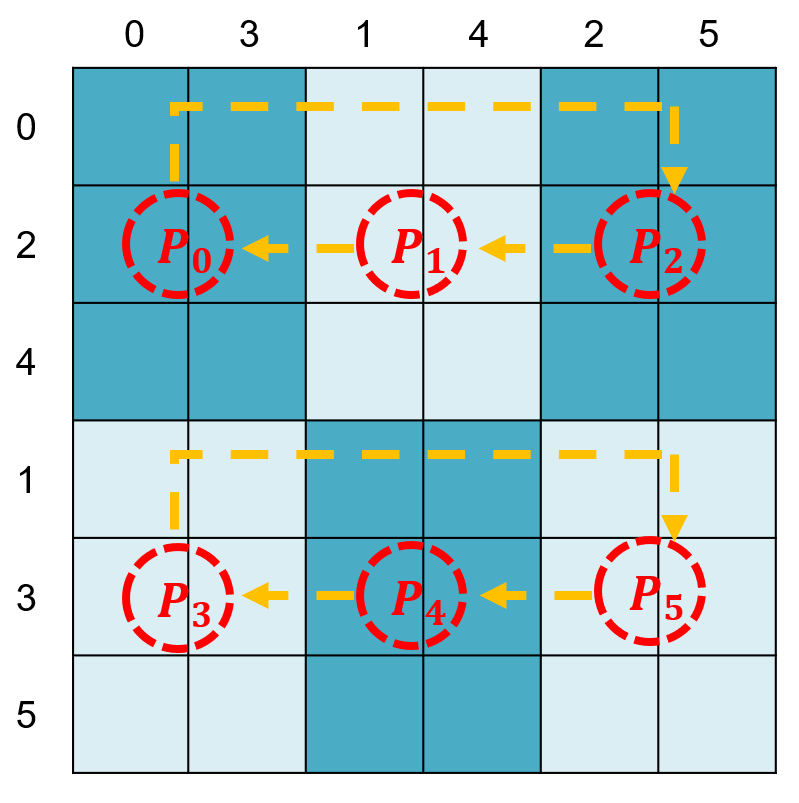}
\label{fig:process}}
\caption{A matrix with $6\times 6$ blocks is distributed over a $2\times 3$ process grid.}
\label{fig:block-cyclic}
\end{figure}

Suppose the matrix $A$ has $M$ block rows and $K$ block columns, and the matrix
$B$ has $K$ block rows and $N$ block columns. Block $(I,J)$ of $C$ is then computed by
\begin{equation}
\label{eq:MatC}
C(I,J) = \sum_{\ell=0}^{K-1} A(I,\ell)\cdotp B(\ell,J),
\end{equation}
where $I=0,1,\cdots,M-1, J=0,1,\cdots,N-1$.
Cannon and Fox algorithms initially considered only the case of matrices in which
each processor contains a single row or a single column of blocks~\cite{Cannon_MM,fox1987matrix}.
PUMMA considered the matrix multiplication algorithms with BCDD form~\cite{1997summa}.
Fig.~\ref{fig:block-cyclic} shows a $6\times 6$ block matrix stored in a
$2\times 3$ process grid.
It is easy to see that the matrix in the block cyclic form is obtained from
the original matrix by performing row and column block permutations.

Since the matrix $B$ can be represented by its generators,
any submatrices of $B$ can be formed easily.
This fact enables us to treat the submatrices of $A$ in each process as
a whole continuous block, and we only need to construct the proper
submatrices of $B$ correctly.
This is our \textbf{main observation}.
After discovering this fact, the proposal of the PSMMA algorithm becomes very natural
and simple, just following the equation~\eqref{eq:MatC}.
The whole procedure is shown in Algorithm~\ref{alg:psmma}.

\begin{algorithm}[htb]
\label{alg:psmma}
\SetAlgoNoLine
\KwIn{$A\in\mathbb{R}^{m\times k}, B\in \mathbb{R}^{k\times n}$, where $A$ is distributed over a
$p\times q$ process grid, $B$ is a structured matrix and all processes have a copy of its generators;}
\KwOut{$C = A \times B$.}
\begin{enumerate} \setlength{\itemsep}{0pt}
\item Each process constructs the column indexes (\emph{CIndex}) of matrix $B$ based on its process column in the process grid;

\item Set $C=0.$

\item \textbf{do} $\ell=0,q-1$

  \begin{itemize}

  \item[(a)] For each process $(i,j)$, construct the column indexes (\emph{RIndex}) of matrix $A$ based on the process column $\mod(j+\ell,q)$;

  \item[(b)] Calculate the required $B$ subblock $B(\emph{RIndex,CIndex})$, and construct its low-rank approximation (if needed) by using its generators, $B \approx U_B V_B$;

  \item[(c)] Multiply the copied $A$ subblock with the currently residing $B$ subblock: $C = C+(A\cdotp U_B) \cdotp V_B$;

  \item[(d)] Shift matrix $A$ leftward cyclically along each process row;

  \end{itemize}

 \textbf{end do}

  \end{enumerate}
\caption{PSMMA for a structured matrix $B$.}
\end{algorithm}

\begin{figure}[ptbh]
\centering
\includegraphics[width=3.5in,height=3.5in]{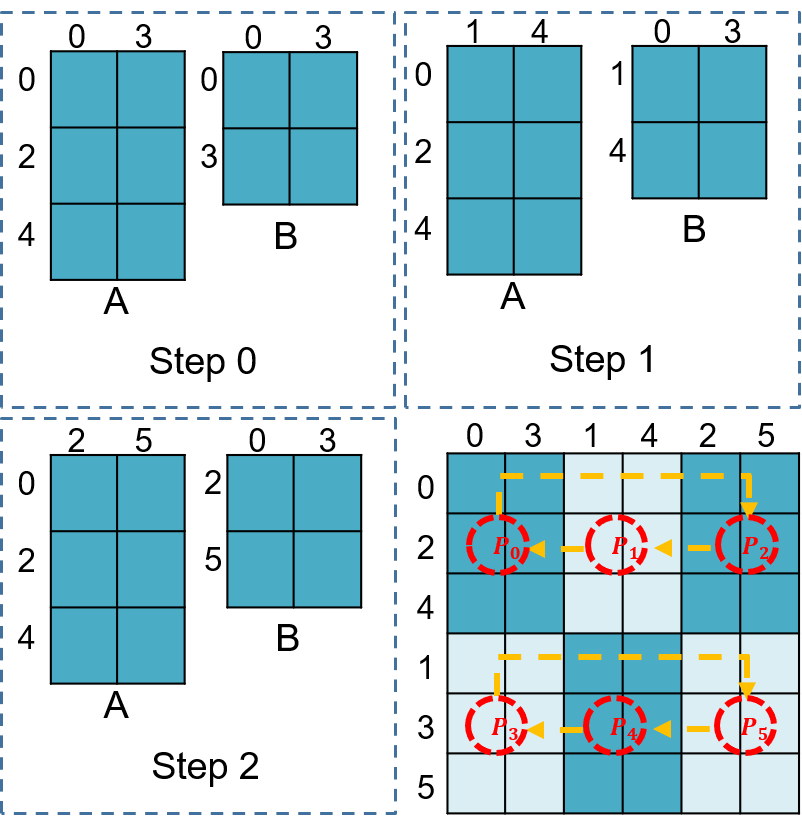}
\caption{The process for computing the submatrices of $C$ located at
process $(0,0)$.}
\label{fig:psmma_C}
\end{figure}

To illustrate the algorithm from the process point of view, we show how the submatrices of $C$ stored at
process $P_0$ (located at position $(0,0)$ of the process grid) are computed for the matrix shown in Fig.
\ref{fig:process}. This consists of three steps, and the process is
depicted in Figure~\ref{fig:psmma_C}.
The column indexes of $B$ are fixed, and its row indexes are determined by the
column indexes of $A$.
After each step, matrix $A$ would be shifted leftward,
and the local matrix $A$ on process $P_0$ is updated
by the matrix on process $P_1$.
As shown in Fig.~\ref{fig:process}, the local matrix $A$ of
process $P_0$ at step $1$ is updated by that of process $P_1$,
which is located at the right of process $P_0$.
After process $P_0$ has received the matrix $A$ from all other processes,
the algorithm stops.

Algorithm~\ref{alg:psmma} works both for block cyclic data distribution and block data
distribution.
It only depends on the distribution of $A$ to determine the row indexes of local matrix $B$.
In step 3(b) of Algorithm~\ref{alg:psmma}, we construct a low rank approximation only
when the local matrix $B$ is probably low rank.
For the tridiagonal DC algorithm in section~\ref{sec:prelim}, matrix $B$ is
a Cauchy-like matrix and we can check whether the intersection of \emph{CIndex} and
\emph{RIndex} is empty or not. If the intersection is empty, the $B$ submatrix is probably
numerically low-rank. Otherwise, the $B$ submatrix is probably full rank.
For Cauchy-like matrices, we use SRRSC discussed in section~\ref{sec:srrsc} to
construct a low-rank approximation to matrix $B$.

\begin{remark}
The PSMMA algorithm introduced in this section is also suitable for other
structured matrices, such as Toeplitz, Vandermonde, and DFT (Discrete Fourier Transform) matrices.
Some results will be shown in our future works.
\end{remark}

\begin{remark}
Only step \emph{3(d)} of Algorithm~\ref{alg:psmma} requires point-to-point communications.
It can be overlapped with other computations if implemented carefully.
During our numerical experiments, we did not implement this technique.
\end{remark}

\subsubsection{Storage form affects the off-diagonally low-rank property}
\label{sec:comments}

For our problem, the eigenvector matrix $\widehat{Q}$ in
equation~\eqref{eq:rankone} is a Cauchy-like matrix, see equation~\eqref{eq:Q_Cauchy}.
It is further off-diagonally low rank, since $\{d_i\}_{i=1}^n$ and $\{\lambda_i\}_{i=1}^n$
are interlacing.
The storage form of matrix $A$ affects the low-rank property of matrix $B$ in Fig.~\ref{fig:psmma_C}.
We use the following example to show the main points.

\begin{figure}[ptbh]
\centering
\subfigure[Block cyclic distribution of matrix $B$ when $N_B=128$.]{
\includegraphics[width=1.6in,height=1.6in]{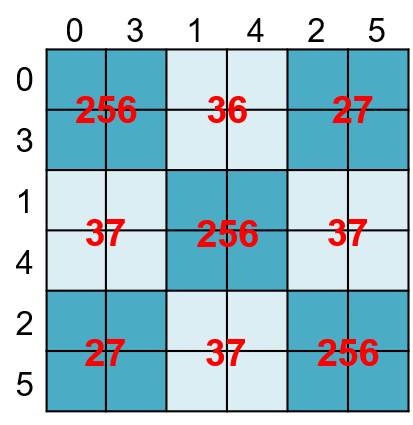}
\label{fig:BlockcyclicB}}
\subfigure[Block distribution of $B$ when $N_B=256$.]{
\includegraphics[width=1.6in,height=1.6in]{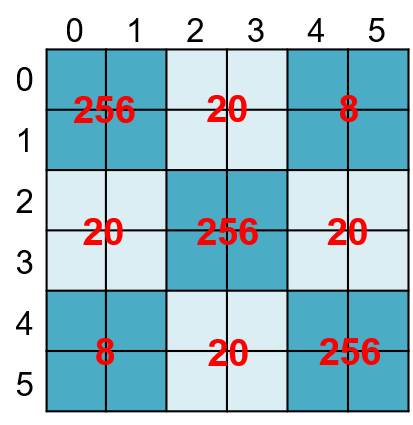}
\label{fig:BlockB}}
\caption{The ranks of $B$ blocks when $B$ is in the BCDD and BDD forms, respectively.}
\label{fig:exam0}
\end{figure}

\textbf{Example 0.}
Assume that matrix $B$ is defined as $B_{ij}=\frac{u_iv_j}{d_i-w_j}$,
where $u_i$ and $v_j$ are random numbers, $d_i=i\cdotp \frac{b-a}{n}$, $w_j=d_j+\frac{b-a}{2*n}$,
$a=1.0, b=9.0$, for $i,j=1,2,\cdots,n.$ It is known that $B$ is a rank-structured Cauchy-like matrix, see~\cite{Shengguo-SIMAX2}.
Let $n=768$ and assume $B$ is distributed over a $3\times 3$ process grid.
Suppose that $N_B=128$ (the parameter of block size for distribution),
we get the BCDD form of matrix $B$, as shown in Fig.~\ref{fig:BlockcyclicB}.
By choosing $N_B=256$, we get the BDD form of $B$ in Fig.~\ref{fig:BlockB}.

Fig.~\ref{fig:exam0} shows the block partitions of matrix $B$ when choosing different $N_B$.
The numbers in Fig.~\ref{fig:exam0} are the ranks of the corresponding blocks.
Fig.~\ref{fig:BlockcyclicB} shows the ranks of blocks of $B$ when $B$ is in the BCDD form.
From it we can see the off-diagonal ranks in Fig.~\ref{fig:BlockcyclicB} are
larger than those when $B$ is in the BDD form, which are shown in
Fig.~\ref{fig:BlockB}.
Many flops can be saved when the ranks are small.
When $B$ is initially stored in the BCDD form
with small $N_B$, we can transform it to the BDD form
by using ScaLAPACK routine \texttt{PDGEMR2D}.
We compare these two cases in Example 1 in section~\ref{sec:num}.

\subsubsection{Comparison with other algorithms}
\label{sec:comparison}

Algorithm~\ref{alg:psmma} works on the whole block of local matrices,
like the block partition based algorithms such as Cannon and Fox algorithms.
The advantages of PSMMA over Cannon and Fox algorithms are that
Algorithm~\ref{alg:psmma} also works for matrices of BCDD form and
also works for any rectangular process grids.

Comparing with PUMMA, which works for block cyclic data distribution
and rectangular process grids, Algorithm~\ref{alg:psmma} requires less workspace
and the size of local matrix multiplications is larger than that in PUMMA.
Based on the \emph{LCM} concept, PUMMA needs extra workspace to permute the block
columns of $A$ and block rows of $B$ together, which can be multiplied simultaneously.
Therefore, PUMMA requires roughly as much extra workspace to store another copy of
the local matrices $A$ and $B$, which makes it impractical in real applications, see~\cite{choi1994pumma,choi1998new}.
Furthermore, only partial block columns (rows) can be merged together in PUMMA,  while
PSMMA always multiplies the whole local matrix $A$ with $B$ and
it does not need to permute the block columns of $A$ or the block rows of $B$.

Comparing with SUMMA, which is based on the outer product form of matrix multiplication,
PSMMA can further exploit the low-rank structure of matrix $B$.
It is not easy for SUMMA to exploit this property.

\subsection{Parallel Structured DC Algorithm}
\label{sec:deflation}

The excellent performance of the DC algorithm is partially due to \emph{deflation}~\cite{BNS-Rankone,Cuppen81}, which
happens in two cases.
If the entry $z_i$ of $z$ is negligible or zero, the corresponding $(\lambda_i, \hat{q}_i)$ is already an
eigenpair of $T$.
Similarly, if two eigenvalues in $D$ are identical then one entry of $z$ can be transformed to zero by
applying a sequence of plane rotations.
All the deflated eigenvalues are moved to the end of $D$ by a permutation matrix, and so are the
corresponding eigenvectors.
Then, after deflation, \eqref{eq:T3} reduces to
\begin{equation}
\label{eq:T3-def}
T=Q(GP)\begin{pmatrix}\bar{D}+b_k\bar{z}\bar{z}^T & \\ & \bar{D}_d \end{pmatrix}
(GP)^T Q^T,
\end{equation}
where $G$ is the product of all rotations, $P$ is a permutation matrix, and
$\bar{D}_d$ are the deflated eigenvalues.

According to~\eqref{eq:rankone}, the eigenvectors of $T$ are computed as
\begin{equation}
\label{eq:evect}
U=\left[ \begin{pmatrix} Q_1 & \\ & Q_2 \end{pmatrix} GP \right] \begin{pmatrix} \widehat{Q} & \\ & I_d \end{pmatrix}.
\end{equation}
To improve efficiency, Gu~\cite{Gu-thesis} suggested a permutation strategy for reorganizing
the data structure of the orthogonal matrices, which has been used in ScaLAPACK.
The matrix in square brackets is permuted as
$\begin{pmatrix}{Q}_{11} & {Q}_{12} & 0 & {Q}_{14} \\
0 & {Q}_{22} & {Q}_{23} & {Q}_{24} \end{pmatrix}$,
where the first and third block columns contain the eigenvectors that have not been affected
by deflation, the fourth block column contains the deflated eigenvectors, and the second block
column contains the remaining columns.
Then, the computation of $U$ can be done by two parallel matrix-matrix products
(calling \texttt{PDGEMM}) involving parts of $\widehat{Q}$ and
the matrices $\begin{pmatrix} Q_{11} & Q_{12} \end{pmatrix}$,
$\begin{pmatrix} Q_{22} & Q_{23} \end{pmatrix}$.
Another factor that contributes to the excellent performance of DC is that
most operations can take advantage of highly optimized matrix-matrix products.

When there are few deflations, the size of matrix $\widehat{Q}$ in~\eqref{eq:evect} will be large, and
most of the time spent by DC would correspond to the matrix-matrix multiplication in~\eqref{eq:evect}.
Furthermore, it is well-known that matrix $\widehat{Q}$ defined as in~\eqref{eq:rankone} is a Cauchy-like
matrix with off-diagonally low rank property, see~\cite{Gu-eigenvalue,LSG-NLAA}.
Therefore, we simply use the parallel structured matrix-matrix multiplication algorithm
to compute the eigenvector matrix $U$ in~\eqref{eq:evect}.
Since PSMMA requires much fewer floating point operations and
communications than
the plain matrix-matrix multiplication, \texttt{PDGEMM}, this approach
makes the DC algorithm in ScaLAPACK much faster.

\begin{figure}[ptbh]
\centering
\includegraphics[width=3.6in,height=3.8in]{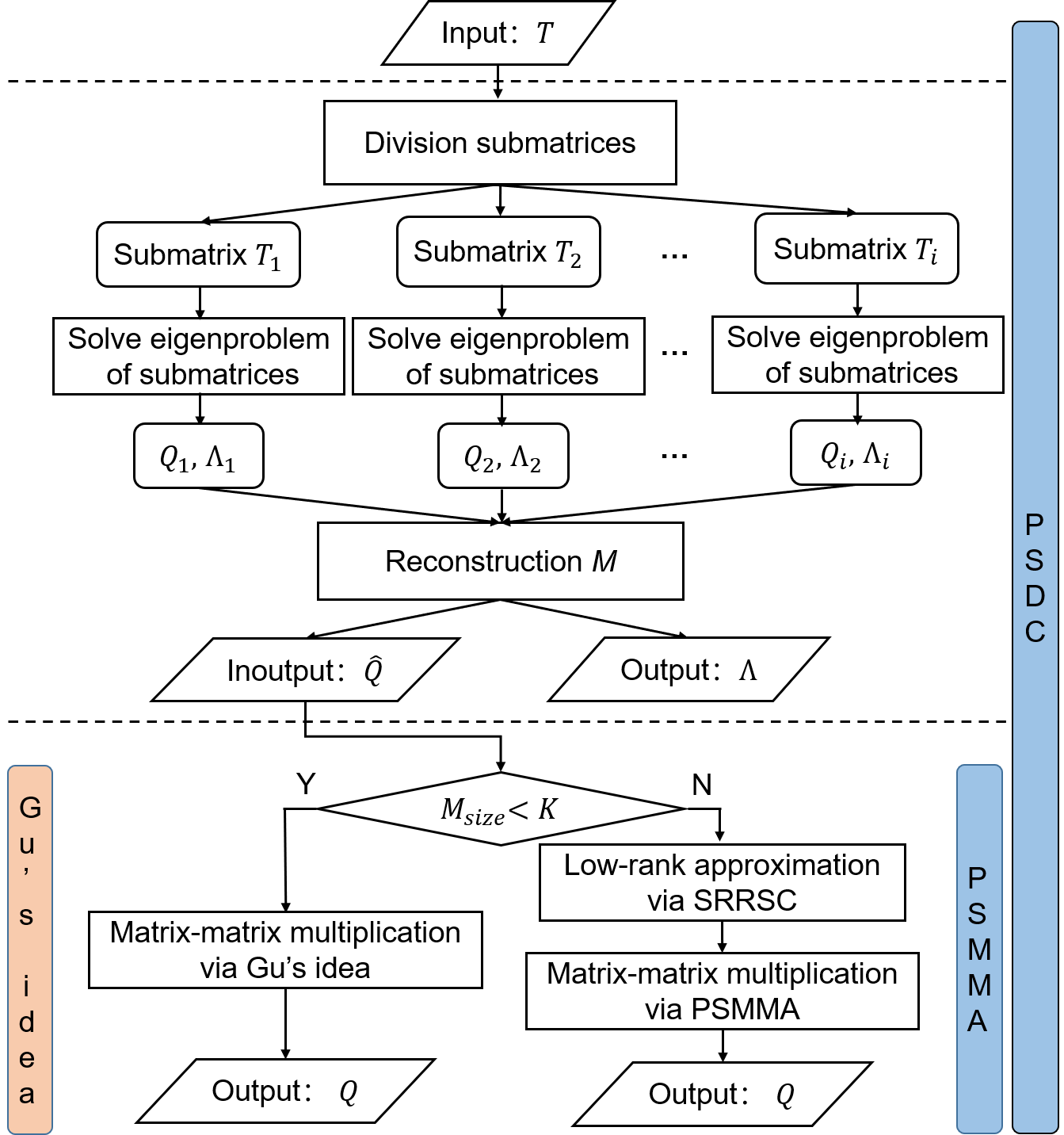}
\caption{The stages of the PSDC method for solving the symmetric tridiagonal eigenvalue problem.}
\label{fig:psdcfig}
\end{figure}

As mentioned before, the central idea is to replace \texttt{PDGEMM} by PSMMA.
The eigenvectors are updated in the ScaLAPACK routine \texttt{PDLAED1}, and
therefore we modify it and call PSMMA in it instead of \texttt{PDGEMM}.
The whole procedure of PSDC accelerated by PSMMA is summarized in Algorithm~\ref{alg:psdc}.
Comparing with the classical DC algorithm (Algorithm~\ref{alg:dc}),
the only difference is that PSMMA is used when the size of matrix $M$ is large.
In Fig.~\ref{fig:psdcfig} the stages of PSDC are graphically represented.

%
%
%
%


\begin{algorithm}[htb]
\label{alg:psdc}
\SetAlgoNoLine
\KwIn{$T\in\mathbb{R}^{n\times n}$}
\KwOut{eigenvalues $\Lambda$, eigenvectors $Q$}
\eIf{ the size of $T$ is small enough}{
apply the QR algorithm and compute $ T = Q \Lambda Q^T $ \;
\Return $Q$ and $\Lambda$\;
}{
form $T= \begin{bmatrix} T_1 & \\ & T_2 \end{bmatrix}+b_k vv^T$\;
\textbf{call} PSDC($T_1,Q_1,\Lambda_1$)\;
\textbf{call} PSDC($T_2,Q_2,\Lambda_2$)\;
form $M = \bar{D}+b_k\bar{z}\bar{z}^T$\ from $Q_1, Q_2, \Lambda_1, \Lambda_2,$ and $v$,
$\bar{D}=\diag(\Lambda_1,\Lambda_2)$ after deflations\;
\eIf{ the size of matrix $M$ is small}{
find eigenvalues $\Lambda$ and eigenvectors $\widehat{Q}$ of $M$\;
use Gu's idea to calculate $ Q= \begin{bmatrix} Q_1 & \\ & Q_2 \end{bmatrix} \widehat{Q}$\;
}{
find eigenvalues $\Lambda$ and eigenvectors $\widehat{Q}$ of $M$\;
use PSMMA (via SRRSC) to calculate $Q= \begin{bmatrix} Q_1 & \\ & Q_2 \end{bmatrix} \widehat{Q}$\;
    }
    \Return $Q$ and $\Lambda$\;
 }
\caption{PSDC($T, Q, \Lambda$) algorithm for computing the eigendecomposition of a symmetric tridiagonal matrix.}
\end{algorithm}

Note that after applying permutations to $Q$ in~\eqref{eq:evect}, matrix $\widehat{Q}$
should also be permuted accordingly.
From the results in~\cite{Gu-eigenvalue,LSG-NLAA,Shengguo-SIMAX2}, we know that
$\widehat{Q}$ is a Cauchy-like matrix and off-diagonally low-rank, the numerical rank is usually
around $50$-$100$.
When combining with PSMMA, we would not use Gu's idea~\cite{Gu-thesis} since permutation may destroy the
off-diagonally low-rank structure of $\widehat{Q}$ in~\eqref{eq:evect}.
We need to modify the ScaLAPACK routine \texttt{PDLAED2}, and only when the size of deflated matrix $\bar{D}$
in~\eqref{eq:T3-def} is large enough, PSMMA would be used, otherwise use Gu's idea.
In section~\ref{sec:num}, we denote the size of $\bar{D}$ by $K$, whose value depends on the matrix as well as the architecture of the particular parallel
computer used, and may be different for different computers.
In Example 2, PSMMA is used when $K\ge 20,000$.

\begin{remark}
To keep the orthogonality of $\widehat{Q}$ (see equation~\eqref{eq:Q_Cauchy}), $d_i-\lambda_j$ must be computed by
\begin{equation}
  \label{eq:d-lambda}
  d_i - \lambda_j =
\begin{cases}
(d_i-d_j)-\gamma_j & \text{ if } i\le j  \\
(d_i -d_{j+1})+\mu_j & \text{ if } i > j
\end{cases},
\end{equation}
where $\gamma_i=\lambda_i-d_i$ (the distance between $\lambda_i$ and $d_i$),
and $\mu_i=d_{i+1}-\lambda_i$ (the distance between $\lambda_i$ and $d_{i+1}$),
which can be returned by calling the LAPACK routine \texttt{DLAED4}.
In our implementation, $\widehat{Q}$ is represented by using \emph{five} generators,
$\{d_i\}, \{\gamma_i\}, \{\mu_i\}, \{u_i\}$ and $\{v_i\}$.
\end{remark}

\section{Experimental results}
\label{sec:num}

All the experimental results are obtained on the Tianhe-2 supercomputer~\cite{Liao-TH2, Liao-HPCG},
located in Guangzhou, China. Each compute node is equipped with two 12-cores Intel Xeon E5-2692 v2 CPUs and our experiments only use CPU cores.
The details of the test platform and environment of compute nodes are shown in Table~\ref{tab:tianhe2a}.
For all these numerical experiments, we only used plain MPI, run $24$ MPI processes per node in principle, and
one process per core.
For example, we used $171$ compute nodes for testing $4096$ processes.

\begin{table}[ptbh]
\caption{The test platform and environment of one node.}
\label{tab:tianhe2a}
\begin{center}%
\begin{tabular}
{|c|c|}\hline
Items  & Values  \\ \hline
2*CPU  & Intel Xeon CPU E5-2692 v2@2.2GHz \\ \hline
Memory size & 64GB (DDR3) \\ \hline
Operating System  & Linux 3.10.0 \\ \hline
Complier &Intel ifort  2013\_sp1.2.144 \\ \hline
Optimization & -O3 -mavx
\\ \hline
\end{tabular}
\end{center}
\end{table}

\textbf{Example 1.} Assume that $A\in \mathbb{R}^{n\times n}$ is a random matrix and
$B$ is defined as $B_{ij}=\frac{u_iv_j}{d_i-w_j}$,
where $u_i$ and $v_j$ are random numbers, $d_i=i\cdotp \frac{b-a}{n}$, $w_j=d_j+\frac{b-a}{2*n}$,
$a=1.0, b=9.0$, for $i,j=1,\cdots,n.$ It is known that $B$ is a rank-structured
Cauchy-like matrix, see~\cite{Shengguo-SIMAX2,LSG-NLAA}.
We compute $C=A\times B$. Let $n=8192$, $16384$ and $32768$, respectively, and choose
different number of processes, $N_P=16$, $64$, $256$, $1024$ and $4096$, to compare
\texttt{PDGEMM} with PSMMA.
To avoid performance variance during multiple executions,
we evaluated the performance of \texttt{PDGEMM} and PSMMA twice in the same program and
called that program three times, and chose the best results among these six executions.
Our codes will be released on Github (available at https://github.com/shengguolsg/PSMMA).

It is well-known that the performance of \texttt{PDGEMM} depends on the block size $N_B$.
We tested the performances of \texttt{PDGEMM} by choosing $N_B=64$, $128$ and $256$, and
we found that their differences are very small.
But they are better than choosing $N_B \le 32$.
Therefore, we chose $N_B=64$ and $n/\sqrt{N_P}$,
which corresponds to the BCDD form and BDD form, respectively.
Note that a large $N_B$ is better for PSMMA since the
ranks of off-diagonal blocks after permutations may be smaller,
see Table~\ref{tab:rank}.

As shown in section~\ref{sec:comments}, we can redistribute matrix $A$ from BCDD to
BDD to exploit the off-diagonal low-rank property of matrix $B$.
In this example, we tested four versions of PSMMA, which are
\begin{itemize}
  \item \textsc{PSMMA\_BCDD}: $N_B=64$ and with low-rank approximation;
  \item \textsc{PSMMA\_BDD}:  $N_B=n/\sqrt{N_P}$ and with low-rank approximation;
  \item \text{PSMMA\_WRedist}: $N_B=64$ and with data redistribution and low-rank approximation
                               (matrix $A$ is transformed from BCDD to BDD with $N_B=n/\sqrt{N_P}$ and then back);
  \item \text{PSMMA\_NLowrank}: $N_B=64$ and without low-rank approximation;
\end{itemize}

\begin{figure}[ptbh]
\centering
\subfigure[Dimension $n=8192$.]{
\includegraphics[width=3.5in,height=2.5in]{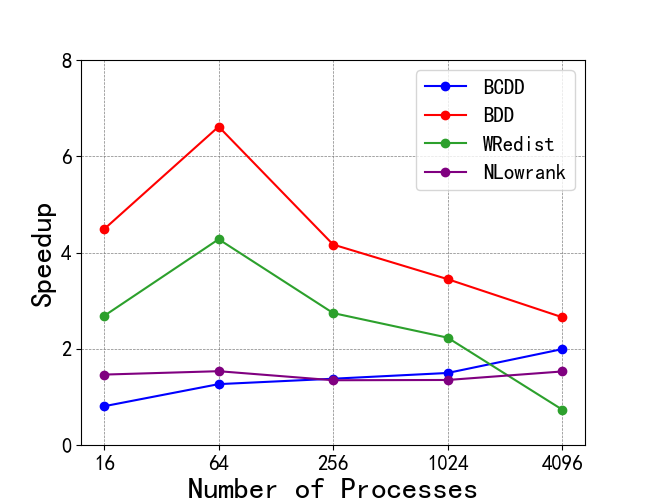}
\label{fig:speedA}}
\subfigure[Dimension $n=16384$.]{
\includegraphics[width=3.5in,height=2.5in]{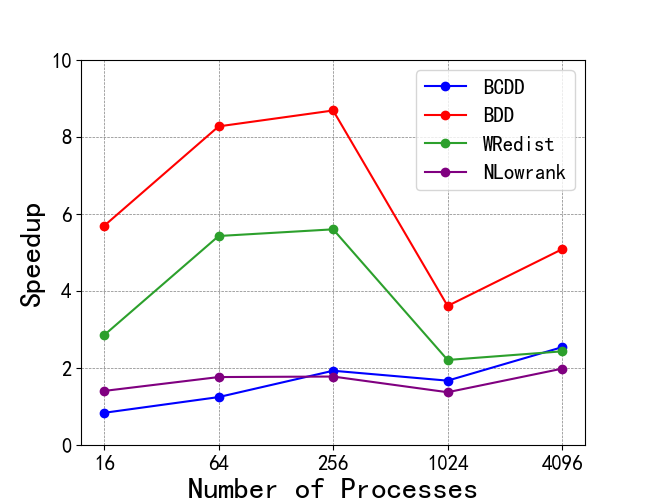}
\label{fig:speedB}}
\subfigure[Dimension $n=32768$.]{
\includegraphics[width=3.5in,height=2.5in]{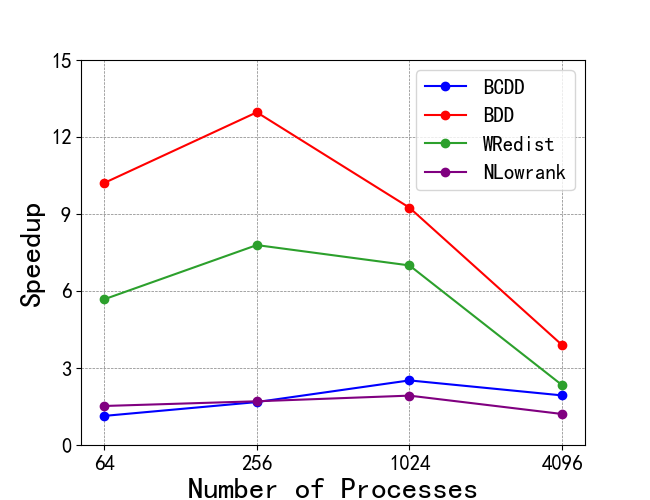}
\label{fig:speedC}}
\caption{The speedup of PSMMA over \texttt{PDGEMM}.}
\label{fig:smmm}
\end{figure}

The speedups of PSMMA over \texttt{PDGEMM} are shown in Fig.~\ref{fig:speedA},
\ref{fig:speedB} and \ref{fig:speedC} with dimensions $n=8192$, $16384$, $32768$,
respectively.
From the results, we can see that PSMMA\_BCDD is always faster than \texttt{PDGEMM}
except for $N_P=16$.
It is because the ranks of $B$ blocks are very large when
using the BCDD form, and most of the time is spent in
computing the low-rank approximations.
Table~\ref{tab:rank} shows the ranks of the off-diagonal blocks
in the first block column, see Fig.~\ref{fig:exam0} for the structure of
matrix $B$.
Without using low-rank approximation, PSMMA can be faster than \texttt{PDGEMM}.
It is interesting to see that PSMMA\_NLowrank is always faster than
\texttt{PDGEMM} for these three matrices.
Note that PSMMA\_NLowrank requires more floating point operations than \texttt{PDGEMM}
since it needs to construct the local $B$ submatrices.
However, PSMMA\_NLowrank requires fewer communications than \texttt{PDGEMM}.

\begin{figure}[ptbh]
\centering
\includegraphics[width=3.5in,height=3.0in]{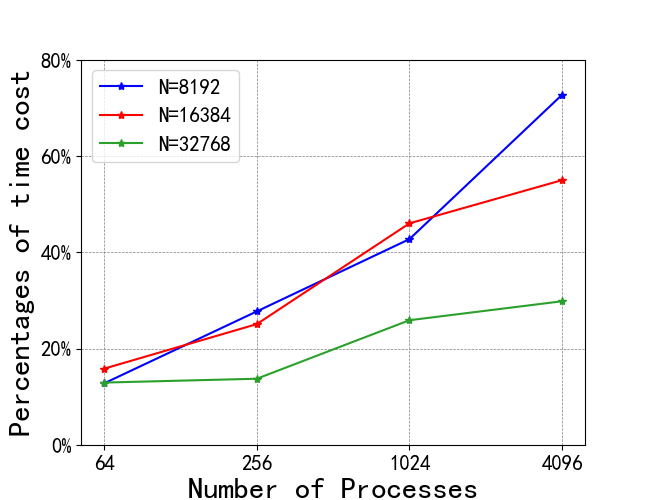}
\caption{The percentages of time required by data redistribution.}
\label{fig:percentage}
\end{figure}

PSMMA\_WRedist is only slower than \texttt{PDGEMM} when
the size of the matrix is small $(n=8192)$ and the number of processes is large.
From the last row of Table~\ref{tab:rank}, we can see that
the ranks of off-diagonal submatrices in BDD form are much smaller than the sizes $(4096)$ of submatrices.
PSMMA\_WRedist is generally faster than both PSMMA\_BCDD and PSMMA\_NLowrank.
The disadvantage of PSMMA\_WRedist is that it requires to perform data redistribution (communication).
When the number of processes is large, the data redistribution represents a large portion of the overall time.
Fig.~\ref{fig:percentage} shows the percentages of transforming matrix $A$ from BCDD to BDD
and transforming it back.
It takes over $70\%$ of total time when $n$ is small and $N_P$ is large.

PSMMA\_BDD is always the best and can be more than ten times faster than \texttt{PDGEMM}.
This is because it assumes that matrix $A$ is initially stored in BDD form and
the low-rank property of matrix $B$ is not destroyed either.

It is difficult to know exactly why the speedups of PSMMA\_WRedist and PSMMA\_BDD
over \texttt{PDGEMM} firstly increase and then decrease.
It depends on the properties of these two algorithms.
Comparing with \texttt{PDGEMM}, PSMMA\_BDD and PSMMA\_WRedist save both computations and communications.
When the number of processes increases, these two contributions make the speedups first increase.
As the number of processes increases, the size of the local submatrix located on each process decreases,
and therefore the percentage of floating point operations
saved from using low-rank approximations decreases.
Since PSMMA\_BDD does not save many floating point operations compared to \texttt{PDGEMM},
the speedups decrease when using more processes.
For PSMMA\_WRedist, the cost of data redistribution also increases as the number
of processes increases.
Meanwhile, as the number of processes grows, the percentage of saved
communication by PSMMA\_BDD and PSMMA\_WRedist increases.
Therefore, PSMMA\_BDD can be always faster than \texttt{PDGEMM} since it not only reduces
computations but also communications.
It is better to use PSMMA\_NLowrank instead of PSMMA\_WRedist when
the number of processes is very large.

\begin{remark}
We can adaptively choose a particular PSMMA algorithm based on the size of matrix and the number
of processes.
It is better to use PSMMA\_WRedist when the number of processes is small, and
use PSMMA\_BCDD or PSMMA\_NLowrank when the number of processes is large and/or
the size of matrix is small.
It is always best to use PSMMA\_BDD when it is available and matrix $B$ is off-diagonally
low rank.
\end{remark}


\begin{table}[ptbh]
\caption{The ranks of off-diagonal blocks of $B$ in the first block column when
stored on $4\times 4$ processes in the BCDD form.
Each block is a $4096\times 4096$ submatrix.}
\label{tab:rank}
\begin{center}
\begin{tabular}[c]{|c|ccc|} \hline
$N_B$ & $B(2,1)$ & $B(3,1)$ & $B(4,1)$ \\ \hline
$64$ & $1260$ & $892$ & $1252$ \\ \hline
$128$ & $711$ & $445$ & $699$ \\ \hline
$256$ & $401$ & $221$ & $390$ \\ \hline
$4096$ & $34$ & $11$ & $9$ \\ \hline
\end{tabular}
\end{center}
\end{table}

\textbf{Example 2.} We use some 'difficult' matrices~\cite{A880} for the DC algorithm,
for which few or no eigenvalues are deflated. Examples include
the Clement-type, Hermite-type and Toeplitz-type matrices, which are defined as follows.

The Clement-type matrix~\cite{A880} is given by
\begin{equation*}
\small
  \label{eq:Clement-Tri}
  T=\text{tridiag}
  \begin{pmatrix}
    &\sqrt{n} & & \sqrt{2(n-1)} & &  & \sqrt{n\cdotp 1} & \\
    0 & & 0 & & \ldots & 0 & & 0 \\
    &\sqrt{n} & & \sqrt{2(n-1)} &  & & \sqrt{n\cdotp 1} & \\
  \end{pmatrix},
\end{equation*}
where the off-diagonal entries are $\sqrt{i(n+1-i)}, i=1,\ldots,n$.

The Hermite-type matrix is given as~\cite{A880},
\begin{equation*}
\small
  \label{eq:Hermite-Tri}
  T=\text{tridiag}
  \begin{pmatrix}
    &\sqrt{1} & & \sqrt{2} & & & \sqrt{n-1} & \\
    0 & & 0 & & \ldots & 0 & & 0 \\
    &\sqrt{1} & & \sqrt{2} & & & \sqrt{n-1} & \\
  \end{pmatrix}.
\end{equation*}

The Toeplitz-type matrix is defined as~\cite{A880},
\begin{equation*}
\small
  \label{eq:Laguerre-Tri}
  T=\text{tridiag}
  \begin{pmatrix}
    &1 & & 1 & & 1 & & 1 & \\
    2 & & 2 & & \ldots & & 2 & & 2 \\
    &1 & & 1 & & 1 & & 1 & \\
  \end{pmatrix}.
\end{equation*}

\begin{figure}[ptbh]
\centering
\subfigure[The strong scaling of PSDC.]{
\includegraphics[width=3.5in,height=3.0in]{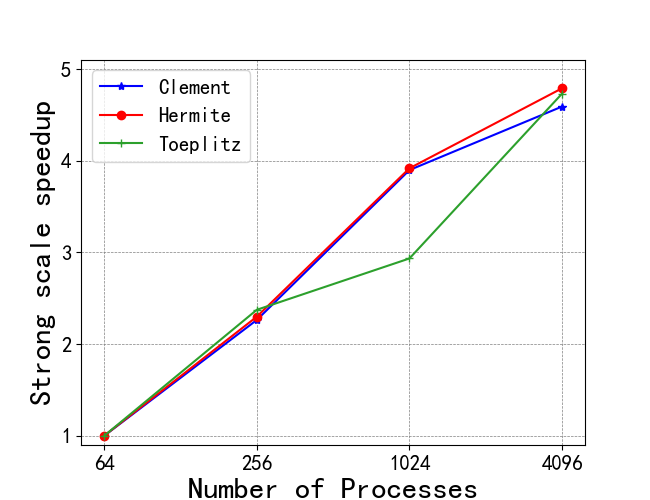}
\label{fig:strong}}
\subfigure[The speedup of PSDC over ScaLAPACK.]{
\includegraphics[width=3.5in,height=3.0in]{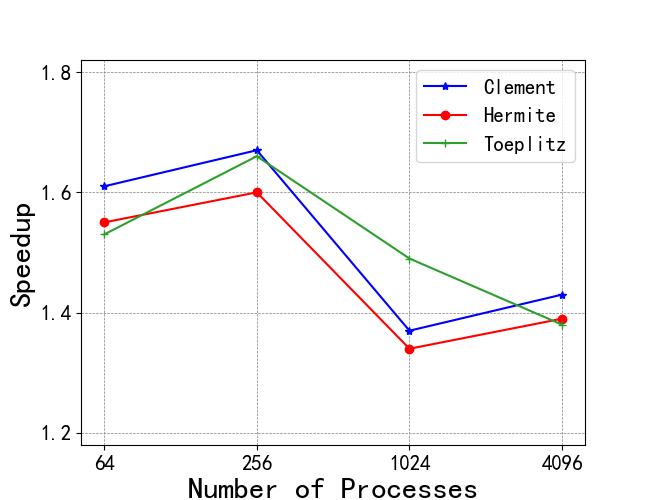}
\label{fig:padc-scal}}
\caption{The results for the matrices of Example 2.}
\label{fig:padc-ex3}
\end{figure}

For the results of strong scaling, we let the dimension $n$ be $30,000$, and use
rank-structured techniques only when the size of the secular equation is larger than $K=20,000$.
We used PSMMA\_WRedist and chose $N_B=64.$
The results for strong scaling of PSDC are shown in Fig.~\ref{fig:strong}.
The speedups of PSDC over ScaLAPACK are reported in Fig.~\ref{fig:padc-scal}.
We can see that PSDC is about $1.4$x--$1.6$x times faster than \texttt{PDSTEDC} in ScaLAPACK for all cases.
Because of deflations, the performances of these three matrices can be
different even though they have the same dimensions.


The orthogonality of the eigenvectors computed by PSDC is
in the same order as those by ScaLAPACK,
as shown in Table~\ref{tab:Ex3-orth}.
The orthogonality of matrix $Q$ is defined as $\|I-QQ^T\|_{\max}$,
where $\| \cdot \|_{\max}$ is the maximum absolute value of entries of $(\cdot)$.
We confirm that the residuals of eigenpairs computed by PSDC are in the same order
as those computed by ScaLAPACK though the results are not included here.

\begin{table}[ptbh]
\caption{The orthogonality of the computed eigenvectors by PSDC.}
\label{tab:Ex3-orth}
\begin{center}%
\begin{tabular}
[c]{|c|cccc|}\hline
\multirow{2}{*}{Matrix}  & \multicolumn{4}{c|}{Number of Processes} \\ \cline{2-5}
  & $64$ & $256$& $1024$ & $4096$  \\ \hline \hline
Clement  & $3.02e$-$14$ & $3.73e$-$14$ & $3.80e$-$14$ & $3.65e$-$14$    \\
Hermite  & $2.49e$-$14$ & $2.75e$-$14$ & $2.96e$-$14$ & $3.01e$-$14$    \\
Toeplitz & $2.88e$-$14$ & $3.01e$-$14$ & $3.03e$-$14$ & $2.97e$-$14$   \\ \hline
\end{tabular}
\end{center}
\end{table}

\begin{figure}[ptbh]
\centering
\includegraphics[width=3.5in,height=3.0in]{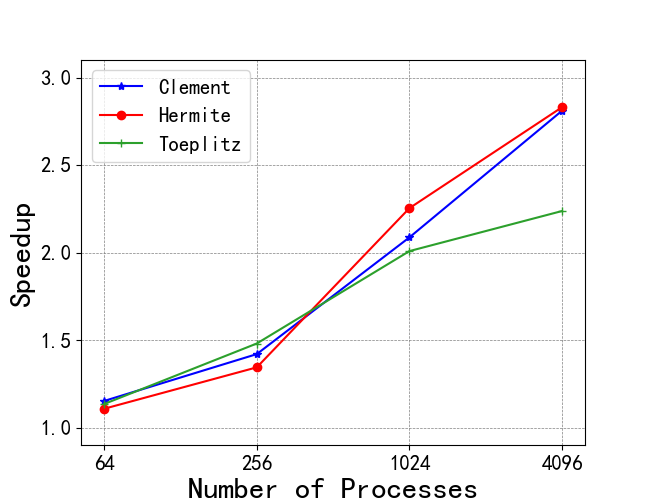}


\caption{The speedup of PSDC over PHDC.}
\label{fig:strumpack}
\end{figure}

Furthermore, we compare PSDC with PHDC which was introduced in~\cite{li2018efficient}
and used STRUMPACK to accelerate the matrix-matrix multiplications.
The results are shown in Fig.~\ref{fig:strumpack}.
For these three matrices, PSDC is much faster than PHDC when using many processes.
It is better to use STRUMPACK when using few processes since HSS-based multiplications
can save more floating point operations than BLR-based multiplications.

\subsection{Results from real applications}

In this subsection, we use three matrices that come from real applications
to test PSDC. One comes from the spherical harmonic transform (SHT)~\cite{Tygert-SHT2}, which
has been used in~\cite{li2018efficient}.
One symmetric tridiagonal matrix is defined as follows, which will be denoted by SHT,
\begin{equation}
A_{jk}=\begin{cases} c_{m+2j-2}, & k=j-1 \\ d_{m+2j}, & k=j \\ c_{m+2j}, & k=j+1 \\ 0, & \mathrm{otherwise}, \end{cases}
\end{equation}
for $j, k=0,1,\ldots,n-1$, where $\xi_l=l-m$,
\begin{equation*}
c_l=\sqrt{\frac{(\xi_l+1)(\xi_l+2)(l+m+1)(l+m+2)}{(2l+1)(2l+3)^2(2l+5)}},
\end{equation*}
\begin{equation*}
d_l=\frac{2l(l+1)-2m^2-1}{(2l-1)(2l+3)},
\end{equation*}
for $l=m, m+1, m+2,\ldots.$
We assume the dimension of this matrix is $n=30,000$ and $m=n$.

The other two are sparse matrices obtained from the
SuiteSparse matrix collection~\cite{TimDavis-Matrix}, called SiO
and Si5H12.
We first reduce each matrix into its tridiagonal form by calling ScaLAPACK
routines and then call PSDC to compute its eigendecomposition.
It is also the general process for computing the eigenvalue decomposition of
any symmetric (sparse) matrices.
These matrices are real and symmetric and their dimensions are
$n=33,401$ and $19,896$, respectively.

\textbf{Example 3.} We use matrices SHT, SiO and
Si5H12 to compare PSDC with \texttt{PDSTEDC}. In this example, we
use the rank-structured techniques, i.e. calling PSMMA,
whenever the size of the secular equation is larger than $K=15,000$,
since the largest $K$ for matrix Si5H12 is $15,489$
when $N_B=64$.
The speedups of PSDC over \texttt{PDSTEDC} are reported in Table~\ref{tab:realapp}.
The backward errors of the computed eigenpairs are also included
in the third column,
which are computed as
\begin{equation}
Residual = \frac{\|A-Q\Lambda Q^*\|_c}{\|A\|_2},
\end{equation}
where $Q\in \mathbb{R}^{n\times n}$ is orthogonal,
$\Lambda\in \mathbb{R}^{n\times n}$ is diagonal with the eigenvalues
as its diagonal elements, $\|X\|_c$ denotes the maximum
Frobenius norm of each column of $X$ and $\|X\|_2$ denotes
the 2-norm of $X$, its largest singular value.


\begin{table}[ptbh]
\caption{The speedups of PSDC over \texttt{PDSTEDC} for matrices from real applications.}
\label{tab:realapp}
\begin{center}%
\begin{tabular}
[c]{|c|c|c|cccc|}\hline
\multirow{2}{*}{Matrix} &\multirow{2}{*}{$K$} & \multirow{2}{*}{Residual} & \multicolumn{4}{c|}{Number of Processes} \\ \cline{4-7}
 &  & & $64$ & $256$ & $1024$ & $4096$ \\ \hline \hline
SHT & $27,136$ & $1.10$e-$14$ & 1.20 & 1.48 & 1.42 & 1.41   \\ \hline
Si5H12 & $15,489$ & $3.54$e-$15$ & 1.27 & 1.16 & 1.15 & 1.05   \\ \hline
SiO & $24,981$ & $1.55$e-$14$ & 1.46 & 1.31 & 1.10 & 1.15   \\ \hline
\end{tabular}
\end{center}
\end{table}

\textbf{Example 4.}
It is known that ELPA (Eigenvalue soLver for Petascale Applications~\cite{Elpa,elpa-library})
has better scalability and is faster than the MKL version of ScaLAPACK.
As opposed to ScaLAPACK, ELPA routines do not rely on BLACS and PBLAS, and
it can overlap the computations with communications, and the
computation is also optimized by using OpenMP and even GPU.
For the tridiagonal eigensolver, ELPA rewrites the DC algorithm and
implements its own matrix-matrix multiplications
and does not use the PBLAS routine \texttt{PDGEMM}.
In contrast, PSDC follows the main procedure of \texttt{PDSTEDC}, and
only uses PSMMA to accelerate the expensive matrix-matrix multiplication part.

\begin{figure}[ptbh]
\centering
\includegraphics[width=3.5in,height=3.0in]{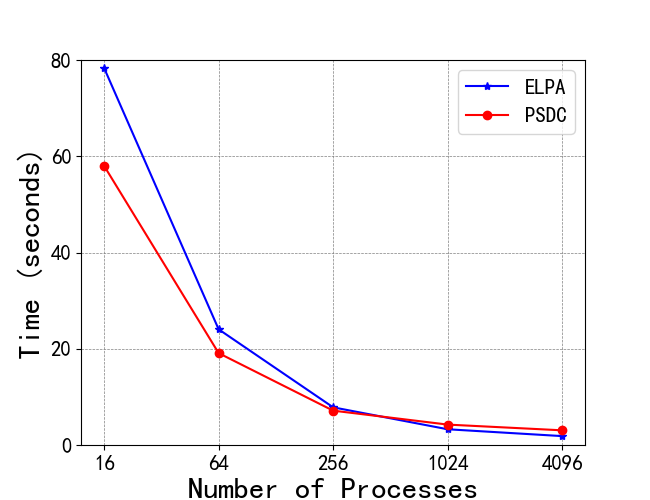}
\caption{The comparison of PSDC with ELPA.}
\label{fig:elpa}
\end{figure}

We compare PSDC with ELPA and find that it competes with ELPA (with version 2018.11.001).
We use the Clement matrix to do experiments.
Fig.~\ref{fig:elpa} shows the execution times of PSDC and ELPA when using
different processes.
It shows that PSDC is faster than ELPA when using few processes,
but PSDC becomes slower when using more than $1024$ processes.
It is because matrix multiplication is no longer the dominant factor
when using many processes.
It is shown in~\cite{li2018efficient} that the percentage of time dedicated to
the matrix multiplications can be less than 10\% of the total time.
The gains of ELPA are obtained from other optimization techniques.

\subsection{Future works}
\label{sec:future}

We discuss some bottlenecks of current implementation and future works in this subsection.
A restriction of our current codes is that PSMMA was only used at the top level
of the DC tree. It should be used at any level as long as the size of the matrix is large.
This will be modified in the near future.
We did not use OpenMP or vectorization to optimize our routines.
The routines for constructing the local submatrices from generators can be
optimized by using vectorization and OpenMP.

Following our current work, there are some interesting research projects to do
in the future.
First, PSMMA can be used to extend the banded DC algorithms
proposed in~\cite{Liao-camwa,Arbenz-bdc} to distributed memory platforms
in a similar manner.
Secondly, the structured matrix-matrix multiplication techniques can be used
in heterogeneous architectures, which can reduce the data movements from CPU to the
accelerators such as GPU\footnote{The authors would like to thank the
referee for pointing out this research direction.}.
We only need to transform the generators to GPUs once instead of many submatrices.
Last but not least, PSMMA can be adapted for Toeplitz, Hankel, DFT (Discrete Fourier Transform)
and other structured matrices~\cite{Bini-Pan}.
Some results will be included in our future works.

\section{Conclusions}
\label{sec:conclusion}

The starting point of this paper is trying to accelerate the tridiagonal DC
algorithm implemented in ScaLAPACK~\cite{Tisseur-DC} for some difficult matrices.
It is known that the main task lies in multiplying a general matrix with a rank-structured
Cauchy-like matrix, see~\cite{Cuppen81,Gu-eigenvalue,LSG-NLAA}.
The main contribution of this paper is that a highly scalable
parallel matrix multiplication algorithm is proposed for rank-structured Cauchy-like matrices,
which fits well for the parallel tridiagonal DC algorithm.

The matrix multiplication problem is known to be very compute
intensive. However, as HPC moves towards exascale computing,
the development of communication-avoiding or communication-decreasing algorithms becomes more and more important.
By taking advantage of the particular structures of Cauchy-like matrices,
we proposed a parallel structured matrix multiplication algorithm
PSMMA, which can reduce both computation and communication costs.
The workflow of PSMMA is similar to PUMMA~\cite{1997summa}
and further exploits the rank-structured property of input matrices.
Experimental results show that PSMMA can be much faster than \texttt{PDGEMM}
for rank-structured Cauchy-like matrices, and the speedups over \texttt{PDGEMM} can
be up to $12.96$.


By combing PSMMA with the parallel tridiagonal DC algorithm, we
propose a parallel structured DC algorithm (PSDC).
For these difficult matrices for which DC deflates very few eigenvalues,
PSDC is always much faster than the classical DC algorithm implemented in ScaLAPACK.
Unlike PHDC which is proposed in~\cite{li2018efficient}, PSDC does not have scalability
problem and it can scale to 4096 processes at least.


%

%
%

\ifCLASSOPTIONcompsoc
  \section*{Acknowledgments}
\else
  \section*{Acknowledgment}
\fi


The authors would like to thank the referees for their valuable comments which greatly
improve the presentation of this paper.
This work is supported by National Natural Science Foundation of China
(No. NNW2019ZT6-B20, NNW2019ZT6-B21, NNW2019ZT5-A10, U1611261, 61872392 and U1811461),
National Key RD Program of China (2018YFB0204303), NSF of Hunan (No. 2019JJ40339),
NSF of NUDT (No. ZK18-03-01), Guangdong Natural Science Foundation (2018B030312002), and
the Program for Guangdong Introducing Innovative and Entrepreneurial Teams under Grant (No. 2016ZT06D211).
Jose E. Roman was supported by the Spanish Agencia Estatal de Investigaci\'{o}n (AEI) under project SLEPc-DA (PID2019-107379RB-I00).

\ifCLASSOPTIONcaptionsoff
  \newpage
\fi

\begin{IEEEbiographynophoto}{Xia Liao}
received the B.S. degree from the College of Computer Science, National University of Defense Technology (NUDT), Changsha, China. She is a PhD candidate in the College of Computer Science at National University of Defense Technology. Her research interests include high performance computing, big data analysis and processing and parallel computing.
\end{IEEEbiographynophoto}

\begin{IEEEbiographynophoto}{Shengguo Li}
received BS, MS and PhD from National University and Defense Technology, Changsha, China,
in computational mathematics. He is currently an assistant professor with the College
of Computer Science, NUDT.
His research interests include numerical linear algebra,
high performance computing, and machine learning.
\end{IEEEbiographynophoto}


\begin{IEEEbiographynophoto}{Yutong Lu}
received the MSc and PhD degrees in computer science from the National University of Defense Technology (NUDT), Changsha, China respectively. She is currently a professor with the School of Data and Computer Science, Sun Yatsen University, Guangzhou, China. She is also the director of National Supercomputer Center in Guangzhou. Her research interests include high performance computing, parallel system management, high-speed communication, distributed file systems, and advanced programming environments with the MPI.
\end{IEEEbiographynophoto}

\begin{IEEEbiographynophoto}{Jose E. Roman}
received the MSc and PhD degrees in computer science from Universitat Politecnica de Valencia (UPV), Spain. He is currently a professor with the School of Computer Science at UPV. His research interests focus mainly on software engineering for large-scale scientific computing, especially in the field of numerical solution of large-scale eigenvalue problems. He is the lead developer of SLEPc, a parallel software library for eigenvalue computations. Other topics of interest are numerical linear algebra, matrix functions, matrix equations, and high performance computing.
\end{IEEEbiographynophoto}



\end{document}